\documentclass[letterpaper, 10 pt, conference]{ieeeconf}  
\IEEEoverridecommandlockouts                              
\usepackage{graphicx} 
\usepackage{amsmath} 
\usepackage{amssymb}  
\usepackage[T1]{fontenc}
\usepackage{comment}
\usepackage{subcaption}
\usepackage{comment}
\usepackage{booktabs}
\usepackage{hyperref}
\DeclareMathOperator*{\argmax}{arg\,max}    

\newtheorem{problem}{Problem}

\newtheorem{proposition}{Proposition}

\usepackage{color} 
\usepackage{xcolor}

\RequirePackage{algorithm}
\RequirePackage{algorithmic}

\definecolor{darkorange}{RGB}{255,140,0}
\definecolor{chocolate}{RGB}{210,105,30}
\definecolor{royalblue}{RGB}{65, 105, 255}
\definecolor{darkgreen}{RGB}{0, 100, 0}
\definecolor{mylimegreen}{RGB}{30, 200, 30}
\definecolor{gray105}{RGB}{105,105,105}
\definecolor{darkmagenta}{RGB}{139,0,139}

\usepackage{tcolorbox}
\usepackage{acronym}

\acrodef{RL}{Reinforcement Learning}
\acrodef{MARL}{Multi-Agent Reinforcement Learning}
\acrodef{MDP}{Markov Decision Process}
\acrodef{MBDP}{Markov Birth-Death Process}
\acrodef{MMPP}{Markov-Modulated Poisson Process}
\acrodefplural{MMPP}[MMPPs]{Markov-Modulated Poisson Processes}
\acrodef{PMF}{Probability Mass Function}
\acrodef{MSS}{Micromobility Sharing System}
\acrodefplural{MSS}[MSSs]{Micromobility Sharing Systems}
\acrodef{MaaS}{Mobility as a Service}
\acrodef{JFI}{Jain Fairness Index}
\acrodef{ML}{Machine Learning}
\title{\LARGE 
	A Fairness-Oriented Reinforcement Learning Approach for the\\ Operation and Control of Shared Micromobility Services
}

\author{Matteo Cederle, Luca Vittorio Piron, Marina Ceccon, Federico Chiariotti,\\ Alessandro Fabris, Marco Fabris, and Gian Antonio Susto 
	\thanks{Corresponding author: M. Ceccon (marina.ceccon@phd.unipd.it).}
	\thanks{ 
		A. Fabris is with the Max Planck Institute for Security and Privacy, 44799 Bochum Universitätsstraße 140, Germany. The remaining authors are with the Department of Information Engineering, University of Padova, 35131 Padua via Gradenigo 6/B, Italy. }
	\thanks{
		This study was partially carried out within the the Italian National Center for Sustainable Mobility (MOST) and received funding from NextGenerationEU (Italian NRRP – CN00000023 - D.D. 1033 17/06/2022 - CUP C93C22002750006).}
}

\begin{document}

\maketitle


\begin{abstract}
	
	As Machine Learning grows in popularity across various fields, equity has become a key focus for the AI community. However, fairness-oriented approaches are still underexplored in smart mobility. Addressing this gap, our study investigates the balance between performance optimization and algorithmic fairness in shared micromobility services providing a novel framework based on Reinforcement Learning.
	Exploiting Q-learning, the proposed methodology achieves equitable outcomes in terms of the Gini index across different areas characterized by their distance from central hubs. Through vehicle rebalancing, the provided scheme maximizes operator performance while ensuring fairness principles for users, reducing iniquity by up to 85\% while only increasing costs by 30\% (w.r.t. applying no equity adjustment). 
	A case study with synthetic data validates our insights and highlights the importance of fairness in urban micromobility (\href{https://github.com/mcederle99/FairMSS.git}{source code}). 
	
	\emph{Index Terms -} Algorithmic Fairness, 
	Q-learning, Reinforcement Learning, Micromobility Sharing Systems, Smart Mobility
\end{abstract}

\section{Introduction}
\label{sec:introduction}

With recent global advances, a growing commitment to focus on how control systems can address large societal challenges has emerged~\cite{alleyne2023control, villa2023fair}. 
In particular, over the past decade, \acp{MSS} have become integral to urban transit~\cite{cheng2021role}, providing last-mile services that complement mass transit and significantly reduce $\text{CO}_{\text{2}}$ emissions. This growth has driven interest in rebalancing techniques~\cite{dell2014bike}, which involve moving shared vehicles to areas of need.
Rebalancing represents 
a significant cost for \ac{MSS} operators, but it is necessary
to consider imbalances in demand patterns and traffic limitations for the trucks that physically transport the vehicles~\cite{chiariotti2020bike}. 

Despite the growth of sharing services, the research community has recently raised a major concern: bikes, scooters, and other micromobility services are more available in wealthier areas, excluding poorer communities~\cite{guan2024shared}, due to higher densities in central areas and lower subscription rates among working-class users~\cite{hosford2018public}, even though the easier access to dockless \acp{MSS}~\cite{brown2021docked} mitigates the problem.
Clearly, unfair systems arising from a lack of attentive policies and profit-oriented management limit accessibility for disadvantaged groups, further marginalizing them and impacting their ability to participate in essential social activities~\cite{chen2020dockless}. For this reason, the concern on seeking equity-based solutions has increasingly gained attention, particularly when opaque learning-based schemes are involved~\cite{caton2024fairness}. 
Specifically, the MSS equity problem is linked to spatial fairness~\cite{soja2009city}, which pursues uniform resource allocation. 
Such a challenge, in turn, translates into balancing the trade-off between minimizing the cost of vehicle placement over densely populated or wealthier areas and adequately distributing the shared vehicles across all neighborhoods, including fairness into the optimization process. 
Our work investigates this trade-off in dockless \acp{MSS}, proposing a \ac{RL} scheme that considers the spatial fairness of the system. The main contributions of this paper are the following.

\begin{itemize}
	\item We propose a simplified fairness-aware \ac{MSS} simulator, by clustering the areas into different categories resting on the proximity to central hub stations.
	\item Through Monte Carlo simulations, we reveal the presence of an inherent trade-off between the MSS performance and the associated fairness level obtained by applying a parametric family of RL-based strategies. 
	\item We analyze the trade-off between spatial fairness and overall cost in \ac{MSS} operation. 
	The proposed method can directly control the balance between fairness, rebalancing costs, and user disservice.    
	\item While the abovementioned works deal with fairness in system planning, to the best of our knowledge, this is the first work on fairness in \ac{MSS} operation and rebalancing. 
\end{itemize}

The remainder of this manuscript unfolds as follows.
Section~\ref{sec:preliminaries} covers the required preliminaries; 
whereas, 
Section~\ref{sec:fair-design} delves into the proposed approach by solving the RL problem through a fairness-oriented design of the reward function.
To support the theoretical findings, Section~\ref{sec:simulations} reports on a 
case study and examines the 
related fairness achievements. 
Lastly, conclusions and future outlooks are sketched in Section~\ref{sec:conclusions}.

\section{System model}
\label{sec:preliminaries}

A dock-based \ac{MSS} is naturally defined as a fully connected graph $\mathcal{G} = (\mathcal{V},\mathcal{E})$, where a node in $\mathcal{V}$ represents a station and $\mathcal{E} = \mathcal{V} \times \mathcal{V}$ denotes the set of connections between each pair of stations. Each node $i \in \mathcal{V}$ is characterized by its current occupancy, i.e., the number of vehicles present at the $i$-th station at time $t$. On the other hand, dockless systems do not have discrete pick-up and drop-off points, as users might leave the shared vehicles anywhere in the service area after their ride. However, the benefits of such an approach, and the extensive literature on docked systems, can be translated to the dockless context by considering service areas instead of stations: the set of nodes $\mathcal{V}$ then becomes a partition of the city map, and each node represents a relatively small area, over which the number of vehicles is counted.

It is vital to observe that \textit{accurately modeling and predicting the dynamics of such networks in their entirety is not a computationally tractable problem for large \ac{MSS} services}, like the ones that we are interested in. We then focus on a stochastic model of an individual service area, considering independent \acp{MMPP}~\cite{ryden1994parameter} for the arrivals and departures, which is consistent with experimental results on large sharing systems~\cite{chiariotti2020bike}. The demand rates vary\footnote{
	The number of requests for vehicles in each area over a short period of time follows a Poisson distribution, with individual events uniformly distributed in time, and the same goes for arrivals. The rate of the Poisson distribution is variable over time and space, following urban mobility trends.
} according to daily, weekly, and seasonal cycles, and are affected by geographic factors as well. 
The vehicle occupancy of the area then follows a left-censored continuous-time \ac{MBDP}~\cite{andronov2011markov}, i.e., a stochastic process in which Poisson events represent either an increase or a decrease of the state by 1, and in which the rate of these events is the outcome of a Markov process with discrete time steps. The left censoring limits the state to nonnegative values: while new arrivals are always
possible (unlike in dock-based systems, in which stations have a maximum capacity), a new departure from the area is impossible if there are no vehicles.
The transition probability from $m$ to $n$ over time $t$ is then approximated by
\begin{equation}\label{eq:skellam}
	P_{m,n}(t)\simeq\begin{cases}
		\sum_{l=m}^{\infty} p_{\text{Sk}}(-l;t,\lambda_a,\lambda_d),& \text{if }n=0;\\
		p_{\text{Sk}}(n-m;t,\lambda_a,\lambda_d), &\text{if }n>0;
	\end{cases}
\end{equation}
in which $p_{\text{Sk}}(n;t,\lambda_a,\lambda_d)$ is the Skellam distribution~\cite{skellam1946frequency}, i.e., the difference of two Poisson random variables:
\begin{equation}\label{eq:skellam_dist}
	p_{\text{Sk}}(n;t,\lambda_a,\lambda_d) \! = \! e^{-t(\lambda_a+\lambda_d)}\!\sqrt{\lambda_a^n\lambda_d^{-n}}I_n\left(2t\sqrt{\lambda_a\lambda_d}\right) \!,
\end{equation}
where $\lambda_a$ and $\lambda_d$ represent the arrival and departure rates, respectively, and $I_n(\cdot)$ is the modified Bessel function of the first kind~\cite{abramowitz1968handbook}. This approximation follows the work in~\cite{chiariotti2020bike}, and its accuracy depends on the frequency with which areas become empty, as its accuracy is decreased by left-censoring: there can never be fewer than $0$ shared vehicles in an area.

In the following, we will consider a system with $V=|\mathcal{V}|$ service areas, which we divide in $M$ categories according to common spatial patterns in U.S. and European cities: in general, central areas tend to see more traffic and have an unbalanced traffic pattern, with more arrivals than departures during the morning rush hours, as commuters tend to go towards commercial areas and large businesses, and more departures during the evening, while residential areas follow an inverted trend with a lower traffic.
Recreational areas such as parks often have yet another pattern, with more trips during the central hours of the day and no rush hour peak, and
recent studies have shown that identifying up to 5 different areas can provide an accurate picture of urban
shared mobility (see, e.g.,~\cite{weinreich2023automatic}).
The rebalancing of each area can then be performed by adding or removing vehicles with a truck, and it is usually a significant cost in \ac{MSS} operation. The overall number of vehicles is often time-varying, as \ac{MSS} operators typically maintain a depot with multiple spare vehicles.

\subsection{MDP formulation}
\label{sec:methodology}
We now illustrate the control approach of this study by modeling the problem as a multi-agent \ac{MDP}~\cite{sutton2018reinforcement} and defining the solution. As a pure \ac{MARL} approach is very complex~\cite{becker2004solving} and may not scale to larger \acp{MSS}, and the interaction between different areas is highly limited, we then devise a way to factorize the problem, optimizing each area separately by solving a single-agent subproblem. A cooperative multi-agent \ac{MDP} is a tuple ${\left\langle \mathcal{S}, \mathcal{A}, \mathcal{P}, \mathcal{R}, \gamma \right\rangle}$, in which $\mathcal{S}$ and $\mathcal{A}$ are two finite and discrete sets, representing the state and action space respectively. In the $N$-agent case, each element of the action space is a vector with $N$ elements, representing the action for each agent.
${\mathcal{P}\left(s, \mathbf{a}, s'\right) = P\left[S(t+1)=s' \mid S(t)=s, A(t)=\mathbf{a}\right]}$ is the state transition probability function, which moves the environment to a new state $s'$ at each iteration, depending on the current state $s$ and the control actions performed by the agents, represented by vector $\mathbf{a}=\begin{bmatrix}
	a_1 &\cdots & a_N
\end{bmatrix}^\top$. Finally, the reward function $\mathcal{R}(s, \mathbf{a}, s'): \mathcal{S} \times \mathcal{A} \times \mathcal{S} \rightarrow \mathbb{R}$ assigns a global reward to all agents, while $\gamma \in [0,1)$ is the discount factor used in the long-term \textit{return}
$G(t) = \sum_{k=0}^{\infty}\gamma^k R(t+k+1)$.
The behavior of the agents is described by their \textit{policy}, i.e., by a function $\pi:\mathcal{S}\to[0,1]^{|\mathcal{A}|}$ that maps each state to a probability of choosing an action vector:
\begin{equation}
	\pi(\mathbf{a}|s) = P[A(t) = \mathbf{a} ~|~ S(t) = s], \hspace{0.3cm} \forall
	\mathbf{a} \in \mathcal{A}.
\end{equation}
We can then define the state value function $v_{\pi}:\mathcal{S}\to\mathbb{R}$, i.e., the expected return\footnote{Notation $\mathbb{E}_{\pi} $ is standard in the RL literature, see \cite{sutton2018reinforcement}.} when the agents follow policy $\pi$:
\begin{equation}
	v_\pi(s) = \mathbb{E}_{\pi} \left[G(t) \mid S(t) = s\right] .  
\end{equation}
A pure \ac{MARL} approach then aims at finding the optimal solution to the multi-agent \ac{MDP}, formally defined as:
\begin{tcolorbox}
	\begin{problem}[
		multi-agent \!MDP]\label{prob:multi}
		\!\!find an optimal policy
		\begin{equation}\label{eq:MAMDP}
			\pi^*=\argmax_{\pi:\mathcal{S}\to\mathcal{A}} v_{\pi}(s),\quad \forall s\in\mathcal{S}.
		\end{equation}
	\end{problem}
\end{tcolorbox}

The elements constituting the system state $s(t)$ are the times of the day (morning or evening), as we consider $2$ rebalancing operations per day, the category of each specific area and the number of vehicles currently available in each service area. The evolution of each area in between rebalancing instants is dictated by the \ac{MMPP} described in \eqref{eq:skellam}, i.e., by means of
the Skellam distribution~\eqref{eq:skellam_dist}. The action space for each agent is designed to be granular enough to offer meaningful choices while keeping an adequate complexity. Actions for each service area include adding or removing up to $30$ vehicles, by increments of $5$.

Finally, the reward function $R(t)$ models the objective of rebalancing operations, i.e., the system operator's profits and operational costs associated with the management of the \ac{MSS}. This economic interest is the combination of various factors. Firstly, the most significant cost in managing \acp{MSS} is represented by rebalancing itself: whenever a truck is dispatched to an area, the operator incurs a cost that is proportional to the centrality of the area. In order to consider the costs of rebalancing different areas and fairness issues between neighborhoods, we partition $\mathcal{V}$ into $\mathbb{P}(\mathcal{V}):=(\mathcal{V}_1,\ldots,\mathcal{V}_M)$, so that $\bigcap_{m=1}^M \mathcal{V}_m = \emptyset$ and $\bigcup_{m=1}^M \mathcal{V}_m = \mathcal{V}$. These $M$ subsets represent the different areas labeled in ascending order from the most peripheral to the most central. 
We also consider a penalty for failures, i.e., whenever a user fails to find a shared vehicle within their service area, which represents the quality of the service, and thus the willingness of users to pay for it. We also include a penalty term for cluttering the sidewalks if there are too many vehicles in the same area: this is a widely discussed issue of \acp{MSS}, which may figure in contracts with city governments, as well as increasing fleet management costs.
Setting 
\begin{equation}\label{eq:rebt}
	\mathrm{reb}_t:= \alpha \sum\nolimits_{m=1}^{M} \left[  \phi(m) \sum\nolimits_{i\in\mathcal{V}_m}[a_{i}(t)]_{*} \right],
\end{equation}
the global reward function is then composed as
\begin{equation}
	\begin{aligned}\label{eq:global_reward}
		R(t) \!=\! 
		- \mathrm{reb}_t
		\!-\sum\nolimits_{i\in\mathcal{V}} f_{i}(t) 
		\!-\!\xi \sum\nolimits_{i\in\mathcal{V}} \!\ell_i(s_{i}^v(t),\mu_{i}(t)),
	\end{aligned}
\end{equation}
where $\alpha,\xi >0$ are constants, 
$[\cdot]_{*}$ is equal to $0$ if the argument is $0$ and $1$ otherwise, and $\phi: \{1,\ldots,M\} \rightarrow [0,1]$ is a strictly decreasing function that satisfies $\phi(1)=1$.
Whenever the action $a_{i}(t)$ is nonzero for the reward function in \eqref{eq:global_reward}, the product $\tilde{\phi}(m):=\alpha\phi(m)$ is subtracted from the total summation; indeed, the latter quantity can be intended as the cost of carrying out a rebalancing operation for the $m$-th area.
Furthermore, the variable $f_{t,i}$ represents the number of failures over the node $i$ during the considered interval, i.e., the number of users who fail to find a shared vehicle in that area. Also, the last term of $R(t)$ accounts for the fact that the injection of further vehicles into the network should be penalized proportionally, due to the clutter and fleet maintenance issues discussed above. Such a cost is modeled proportionally to the sum of every mismatch between the current number\footnote{The quantity $s_{t,i}^v$ is part of the observable state $s_{t,i}$ and it is upper-bounded 
	by $\sigma_i \gg$ $\mu_{t,i}$, $ \forall t\geq 0$, 
	to render the state space finite.} 
of vehicles $s_{t,i}^{v}$ $\in [0,\sigma_i]$ and the  expected demand $\mu_{t,i}$ (until the next rebalancing action) at each node $i$.
To this purpose, we take the function $\ell_i : \mathbb{N} \times \mathbb{N} \rightarrow \mathbb{R} : (z_1,z_2) \mapsto \ell_i(z_1,z_2)$ to be convex and satisfy the following properties for any couple of integers $(z_1,z_2)$: 
$\ell_i(z_1,z_2)= \ell_i(z_2,z_1)$; $\ell_i(z_1,z_2)=0$ if and only if $\left\| z_1 - z_2 \right\| \leq \zeta_{\kappa_i} $, for a fixed\footnote{Constant  $\zeta_{\kappa_i}$ can be interpreted as a fraction of the expected arrivals $\mathrm{\bar{a}}_{\kappa_i}$ in all nodes $i$ of the category $\kappa_i$. Henceforth, we assume that $\zeta_{\kappa_i} = 0.5\mathrm{\bar{a}}_{\kappa_i}$.} $\zeta_{\kappa_i}\geq 0$, with $\kappa_i \in\{1,\ldots,M\}$ being
the index for which $i\in \mathcal{V}_{\kappa_i}$ 
and $\left\| \cdot \right\|$ being any metric.

\subsection{Factorized MDP representation}
\label{subsec:3a}

As we discussed above, factorizing the multi-agent problem into multiple single-agent problems can greatly simplify the task. This is possible due to an independence hypothesis: the \acp{MMPP} representing arrivals and departures in each area are assumed to be independent both from each other and from other areas' processes. Clearly, this assumption does not hold for real systems, as trips begin in an area and end in another a few minutes later, but the approximation error is surprisingly low in large-scale systems~\cite{chiariotti2020bike}: any individual area makes up such a small fraction of the total traffic that local events have negligible effects elsewhere. Actions from one agent then have no effect on the state transitions of others. The problem can then be 
factorized~\cite{becker2004solving} into $V$ single-agent \acp{MDP}, which can be solved individually without losing global optimality.

We then define a single-agent subproblem, involving a single service area, in order to divide the multi-agent \ac{MDP} into more manageable components.
\begin{tcolorbox}
	\begin{problem}[Single-area MDP] \label{prob:single} choosing 
		\begin{equation}\label{eq:local_reward}
			\begin{aligned}
				R_{i}(t) = -\!\alpha\phi(\kappa_i) [a_{i}(t)]_{*}\!-\! f_{i}(t), 
				\!-\! \xi\ell_i (s_{i}^v(t), \mu_{i}(t))
			\end{aligned}
		\end{equation}
		as the reward function, find an optimal policy
		\begin{equation}
			\pi_i^*=\argmax_{\pi_i:\mathcal{S}_i\to\mathcal{A}_i} v_{\pi}(s_i),\ \forall s_i\in\mathcal{S}_i,
		\end{equation}
		where $s_i\in\mathcal{S}_i$ includes the state of the $i$-th service area, as well as the time of the day, and $\mathcal{A}_i$ represents the actions that agent $i$ can take. 
\end{problem}\end{tcolorbox}

\begin{proposition}[Separability of the \ac{MSS} problem \cite{becker2004solving}]\label{prop:separability}
	Given the global reward function in~\eqref{eq:global_reward}, the optimal solution to the multi-agent \ac{MDP} defined by Problem~\ref{prob:multi} is given by the combination of the individual solutions to the agent problems in Problem~\ref{prob:single}.
	The resulting solution then enjoys the convergence properties of single-agent Q-learning.
\end{proposition}

\begin{proof}
	The state is separable, as 
	the transition probability of agent $i$ is only affected by its own action $a_{t,i}$: two components of the state (time of day and area type) evolve deterministically, while the third (available vehicles) follows an independent process in each area. 
	It is also trivial to prove that the global reward function in~\eqref{eq:global_reward} is the sum of 
	each reward function in~\eqref{eq:local_reward}. Distributed Q-learning then converges to the optimal solution for the global problem.
\end{proof}

Separating the global problem into $V$ individual subproblems allows for faster training: the state and action spaces become much smaller, avoiding the curse of dimensionality and granting the quick optimization of  
large \acp{MSS}. Also, the training can be reduced to $M$ agents, since areas in the same class have the same statistics 
and thus follow the same single-agent \ac{MDP}. Each agent $m$ can be trained by exploiting the information coming from all the service areas in $\mathcal{V}_m$.
The agents follow a linearly-annealed $\varepsilon$-greedy policy, which guarantees convergence for the Q-learning algorithm to the optimal solution if the learning rate $\eta$ diverges but its square converges~\cite{szepesvari1997asymptotic}, i.e., $\lim_{T\to\infty}\sum_{t=0}^T\eta_t=\infty$, but $\lim_{T\to\infty}\sum_{t=0}^T\eta_t^2<\infty$.

\section{Fairness-oriented design}\label{sec:fair-design}

Fairness has become a key concern in Machine Learning as algorithms shape societal decisions, with emphasis on avoiding models that reinforce disparities linked to sensitive attributes like race, gender, and socio-economic status~\cite{barocas-hardt-narayanan,du2020fairnessdeeplearningcomputational}. In \acp{MSS}, studies~\cite{duran2020fair, su2024spatial} show biases favoring central, affluent areas, disadvantaging outer regions. This bias often correlates with socio-economic status and ethnicity, increasing discrimination against marginalized groups~\cite{guan2024shared,hosford2018public}. Thus, equity must be integrated into optimization processes~\cite{brown2021docked}, with spatial justice promoting fair resource distribution~\cite{soja2009city}. Previous research has largely focused on assessing the equity of \acp{MSS}. For instance,~\cite{guan2024shared} examines how shared services
like e-scooters
impact transportation equity across income groups in European cities, while \cite{hosford2018public} finds that bike-share access in Canadian cities is generally better in wealthier areas, highlighting a need for expansion in underserved regions. In the US,~\cite{brown2021docked} shows that dockless systems reduce geographic inequalities but have mixed results for racial equity, indicating a need for policy intervention. 
Finally,~\cite{su2024spatial} finds that, while e-scooters in Washington, DC, increase access in disadvantaged neighborhoods, they also exacerbate disparities. Our work is the first to embed fairness metrics directly into the MSS planning process, with a focus on individual-level planning, applying fairness metrics for
system rebalancing operations.

\subsection{Fairness metrics in \acp{MSS}}\label{subsec:fairness}

The meaning of spatial fairness from a user-level perspective is simple: what users see and are affected by is the presence of shared vehicles in their vicinity, as it determines their ability to make use of the system and move across the city. We then consider the probability that a user in a given area, i.e., often a person residing or working in that neighborhood, will be unable to find a vehicle in their immediate vicinity during rush hour. A perfectly fair system would equalize this failure probability all over the system. There is an inherent trade-off with rebalancing efficiency: enforcing fairness constraints necessitates increased movement of rebalancing vehicles to low-demand areas, which are typically more remote. This leads to reduced expected profitability compared to central, high-demand areas.

We thus consider the Gini index as a general fairness metric, following general practice in the field~\cite{duran2020fair}, but apply it to our user-level perspective. The Gini index is a measure of statistical dispersion of a distribution, particularly useful for assessing the equality of access to services within a population. Its values span from 0 to 1, where 0 indicates a perfectly fair system, while 1 indicates high unfairness~\cite{gini1921measurement}. In our context, it is defined as
\begin{equation}\label{eq:gini}
	g(x) = (2M^2\bar{x})^{-1} \sum\nolimits_{m=1}^{M} \sum\nolimits_{n=1}^{M} |x_m - x_n|,
\end{equation}
where $M$ is the number of area categories, $x_m$ denotes the probability of service failure at finding an available vehicle in a given category, and $\bar{x}$ denotes the mean value of $x_m$ over $m=1,\ldots,M$. The Gini index can be computed over categories, instead of individual areas, as neighborhoods belonging to the same category are statistically identical.

\subsection{Fair Reinforcement Learning solution}\label{subsec:fair_reward}

Considering a linear combination of profit and a pure fairness metric as an objective function, instead of a mixed metric that includes both economic incentives and system-level fairness, allows us to control the trade-off between economic and fairness concerns.
As the rebalancing cost $\mathrm{reb}_t$ in \eqref{eq:rebt} is strictly decreasing as we get closer to the city center, while the expected utility of visiting an area is strictly increasing, since the demand is higher and so is the expected number of failures if we do not do so, a profit-oriented algorithm will tend to rebalance central areas much more often, leading to a lower failure probability. However, spatial fairness consideration lead us to aim at equalizing the failure rates between different types of areas.

We can then design a strictly decreasing penalty function $\chi: \{1,\ldots,M\} \rightarrow [-1,1]$, which satisfies $\chi(1)=1$, $\chi(M)=-1$ and $\chi(\lceil m-\bar{m} \rceil) = -\chi(\lfloor \bar{m}+m \rfloor)$ for all $m\in \{1,\ldots,\lceil \bar{m} -1 \rceil\}$, with $\bar{m} = (1+M)/2$, and a fairness weighting parameter $\beta>0$. The product $\tilde{\chi}(m) :=~\beta \chi(m)$ acts as a \textit{temperature}, to measure the degree of importance\footnote{In general, the adjustment of the temperature plays a pivotal role in controlling the delicate balance between optimizing performance metrics \cite{friedler2019comparative}, such as accuracy, and ensuring equity in socio-technical systems.} that is given to central areas with respect to peripheral ones. The definition of $\chi$ may be arbitrarily chosen by system designers, but its strictly decreasing nature introduces a higher penalty for failures in more peripheral areas, counterbalancing the tendency of profit-maximizing algorithms to privilege higher-demand, i.e. central, areas.

We can then add the fairness penalty function to the global reward of our \ac{RL} problem:
\begin{equation}\label{eq:our-reward}
	R^{(f)}(t)=R(t)-\beta \sum\nolimits_{m=1}^{M} \left[  \chi(m) \sum\nolimits_{i\in\mathcal{V}_m}f_{i}(t) \right].
\end{equation}
\vspace{-0.3cm}
\begin{proposition}
	The \ac{MSS} optimization problem using modified reward $R^{(f)}(t)$ is still separable.
\end{proposition}
\begin{proof}
	We can easily find an area-level reward $R_i^{(f)}(t)=R_i(t)-\beta\chi(\kappa_i)f_i(t)$, which only depends on the dynamics of area $i$. The separability of the modified problem is then trivial, since the state space and dynamics of the system do not change.
\end{proof}

\vspace{-0.1cm}
\section{Numerical simulations}
\label{sec:simulations}

\begin{table*}[h]
	\caption{Network characterization and demand hyperparameters}
	\label{tab:params}
	\scriptsize 
	\centering
	\begin{tabular}{lccc}
		\toprule
		Scenario & Number of nodes & Skellam parameters $(\lambda_a,\lambda_d)$ for morning demand & Skellam parameters $(\lambda_a,\lambda_d)$ for evening demand\\
		\midrule
		2 classes & $\{60, 10\}$ & $\{(0.3, 2), (13.8, 7)\}$ & $\{(1.5, 0.3), (10, 13.8)\}$\\
		3 classes & $\{60, 30, 10\}$ & $\{(0.3, 2), (3.3, 1.5), (13.8, 7)\}$ & $\{(1.5, 0.3), (1.5, 3.3), (10, 13.8)\}$\\
		4 classes & $\{60, 40, 20, 10\}$ & $\{(0.3, 2), (0.45, 3), (9.2, 5.1), (13.8, 7)\}$ & $\{(1.5, 0.3), (2.25, 0.45), (6.6, 9.2), (10, 13.8)\}$\\
		5 classes & $\{60, 40, 30, 20, 10\}$ & $\{(0.3, 2), (0.45, 3), (3.3, 1.5),(9.2, 5.1), (13.8, 7)\}$ & $\{(1.5, 0.3), (2.25, 0.45), (1.5, 3.3), (6.6, 9.2), (10, 13.8)\}$\\
		\bottomrule
	\end{tabular}
\end{table*}

\begin{figure*}[htbp]
	\centering
	\begin{subfigure}{0.24\textwidth}
		\centering
		\includegraphics[width=\linewidth]{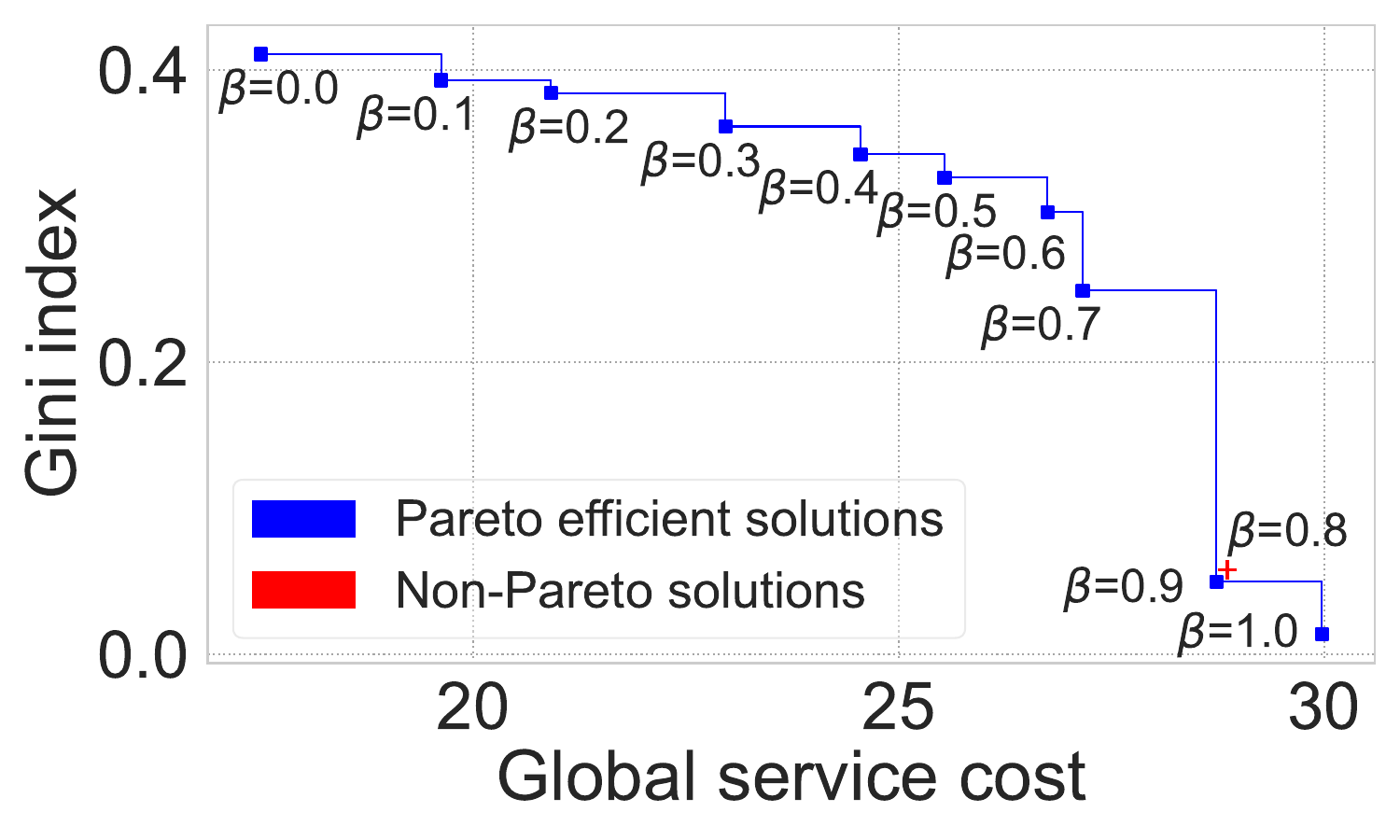}
		\caption{$g(x)$ vs. $\mathcal{C}$ ($M=2$)}
		\label{fig:pareto2}
	\end{subfigure}%
	\begin{subfigure}{0.24\textwidth}
		\centering
		\includegraphics[width=\linewidth]{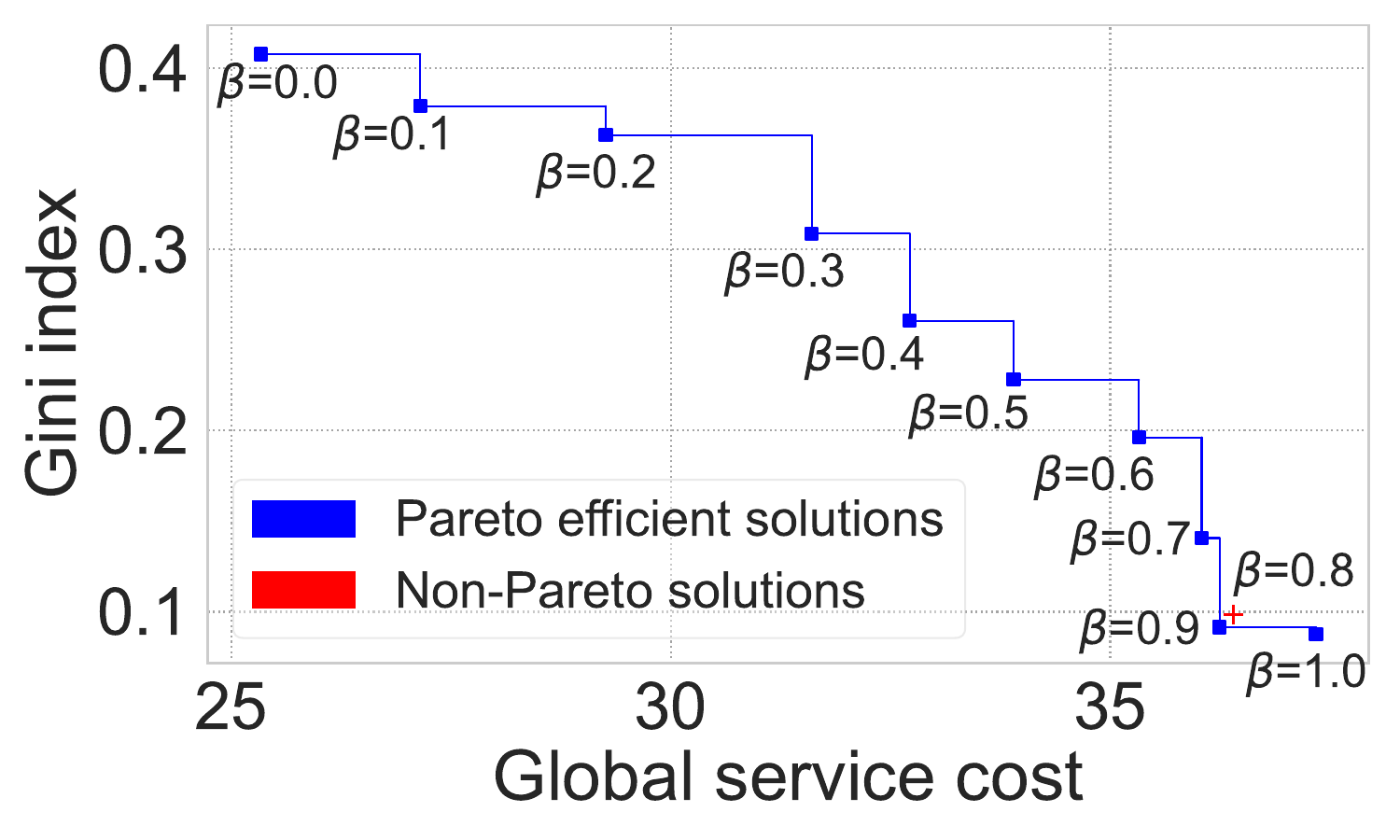}
		\caption{$g(x)$ vs. $\mathcal{C}$ ($M=3$)}
		\label{fig:pareto3}
	\end{subfigure}%
	\begin{subfigure}{0.24\textwidth}
		\centering
		\includegraphics[width=\linewidth]{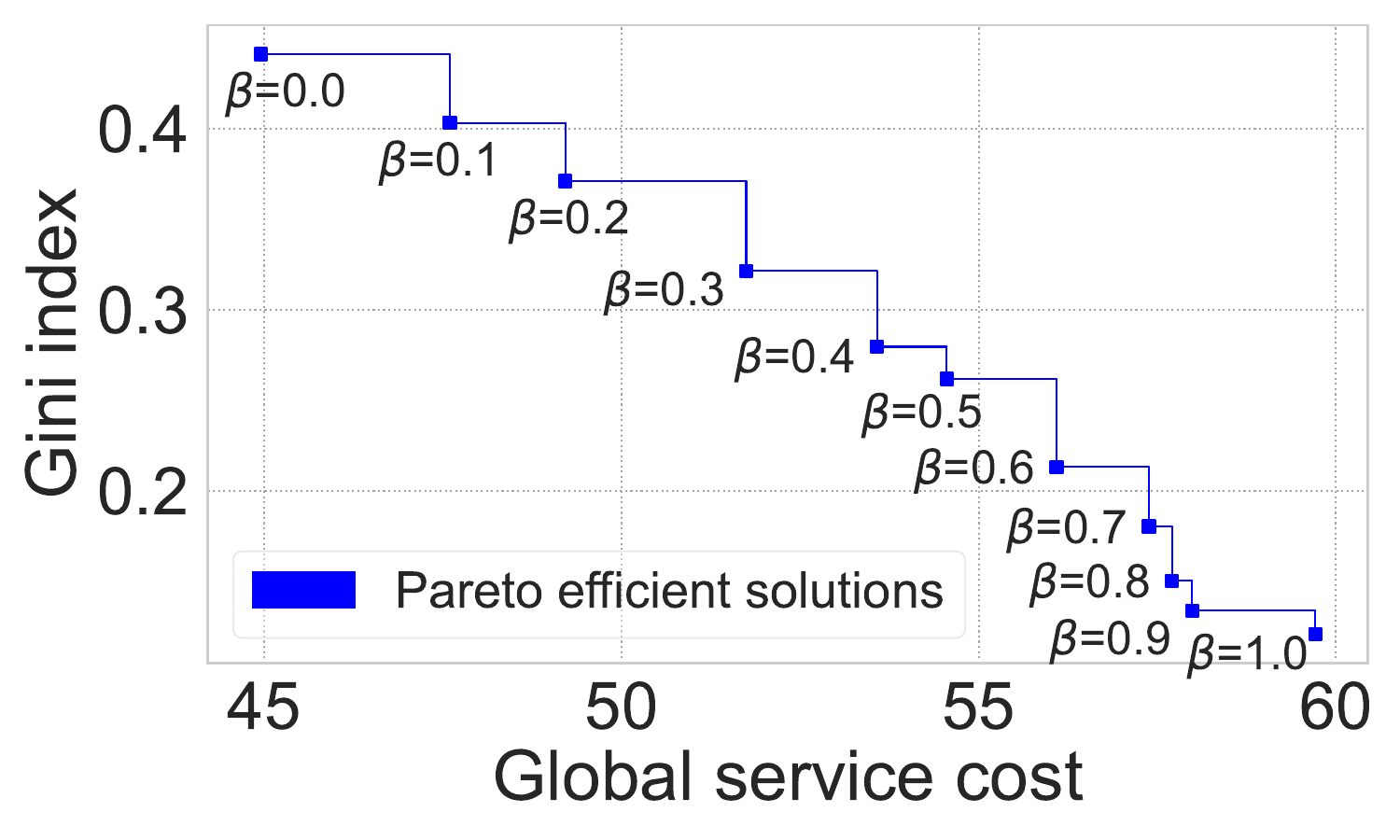}
		\caption{$g(x)$ vs. $\mathcal{C}$ ($M=4$)}
		\label{fig:pareto4}
	\end{subfigure}%
	\begin{subfigure}{0.24\textwidth}
		\centering
		\includegraphics[width=\linewidth]{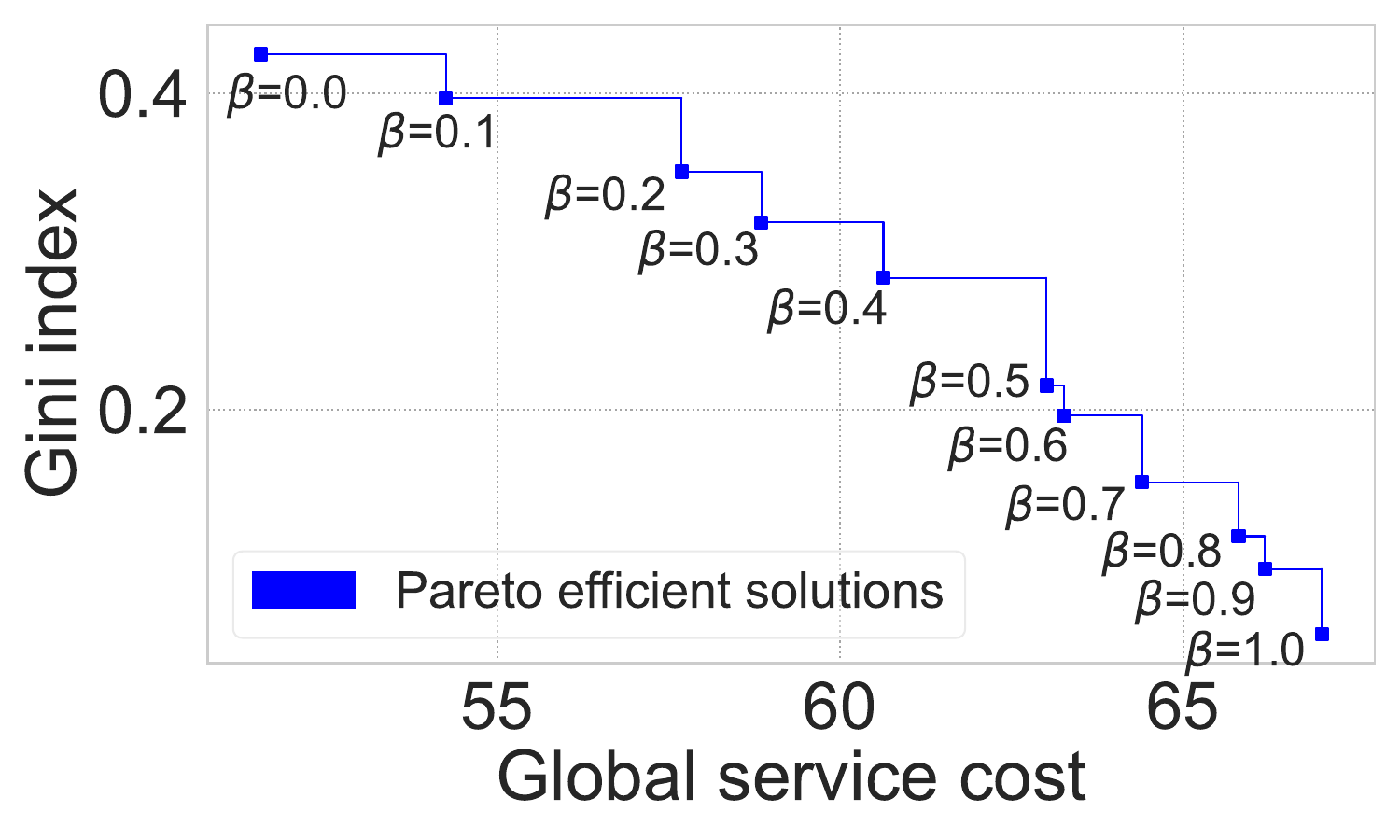}
		\caption{$g(x)$ vs. $\mathcal{C}$ ($M=5$)}
		\label{fig:pareto5}
	\end{subfigure}
	\caption{Pareto fronts for the considered bi-objective optimization problem.
		The cost minimization and the fairness maximization objectives are represented on the x and y axes, respectively. Each mark corresponds to a different value of $\beta$, and the Pareto front (in blue) includes all efficient solutions. The red points correspond to Pareto-inefficient solutions.}
	\label{fig:paretos}
\end{figure*}

We now report on a case study\footnote{Code available at \scriptsize{\url{https://github.com/mcederle99/FairMSS.git}}}
to support the discussed theoretical findings. Next, we shall provide an extensive investigation of different strategies, to demonstrate the trade-off between performance and equity (measured by the Gini index $g(x)$ defined in \eqref{eq:gini}) and find a viable compromise.

In this direction, we have implemented four different experiments, varying the number of categories $M$. On the basis of what pointed out in Section~\ref{sec:preliminaries}, we have examined the cases $M\in\mathcal{M}:=\{2, 3, 4, 5\}$. In each of these scenarios, we have considered a medium-sized micromobility sharing system as an example of dockless \ac{MSS}. The network hyperparameters for each experiment are reported in Table~\ref{tab:params}, where we recall that the categories are ordered from the most peripheral to the most central and follow realistic demand patterns~\cite{weinreich2023automatic}. For each of the experiments the training procedure starts with the service areas being subject to the demand reported in Table~\ref{tab:params}. 
At every hour of the day $t\in\{0,\ldots,23\}$, the number of vehicles present in each area is updated based on the modified MBDP introduced in Section~\ref{sec:preliminaries}.
If at a certain moment a service area is unable to satisfy the demand, i.e. no vehicle is available and there is request for a departure, this is registered as a single failure for that node. The RL agents perform their control actions at 11a.m. and at 11p.m. every day through static rebalancing, as described in Section \ref{sec:introduction}. The training phase for each strategy is run through $T=10^5$ days and evaluated over $E=10^2$ days.  
The learning rate and epsilon decay are set to $0.01$ and $8.25\cdot 10^{-7}$ respectively; while $\gamma:=0.9$, $\alpha:=20$, $\xi:=0.3$ are chosen. 
Finally, $\chi$ takes values from the array $y_\chi := [1,.5,.4,-.5,-1]$ according to its characterization\footnote{The values taken by the functions $\phi$ and $\chi$ generally depend on the distances between different zones and on the regulations established by institutions imposing fairness constraints, respectively. We leave the rigorous investigation of this aspect as future work. as $M$ varies in $\mathcal{M}$; $\phi$~takes values from the array $y_\phi :=~[1,.8,.4,.3,.1]$, so that $\phi(m)=~y_\phi[k]$ if $m,k$ are such that $\chi(m)=y_\chi[k]$;} and $\ell_i(s_{t,i}^v,\mu_{t,i}) := |s_{t,i}^v-\mu_{t,i}| - \zeta_{\kappa_i} \text{, with } \kappa_i\in\{1, \ldots, M\}$.

With the above setup, we analyze the Pareto fronts of the proposed approaches, considering the trade-off between operational costs and fairness.
Also, we examine the scenario $M=5$ in depth, by providing insights about the fairness and costs trends, as the fairness weighting parameter $\beta$ varies.
\begin{figure}[t]
	\centering
	\begin{subfigure}{0.24\textwidth}
		\centering
		\includegraphics[width=\linewidth]{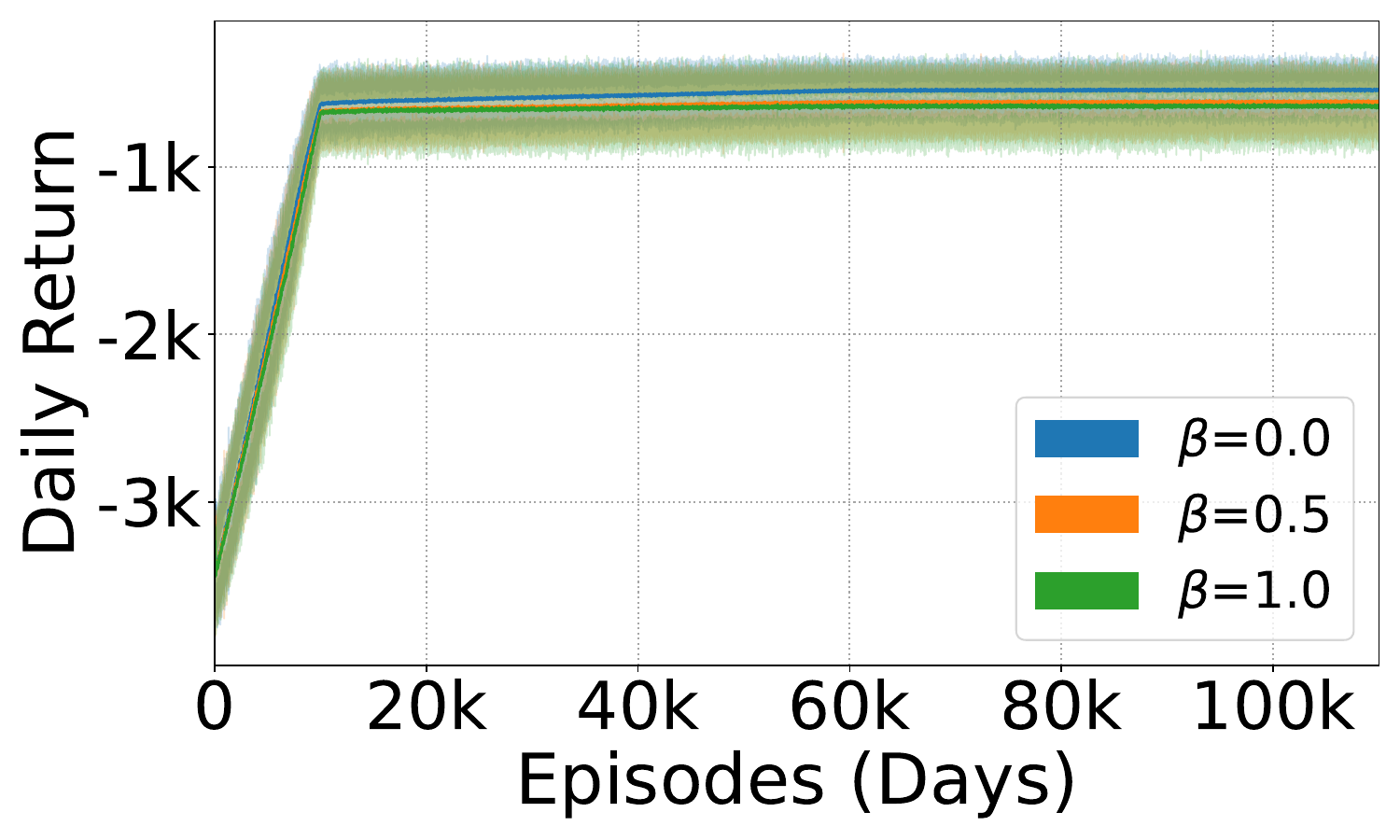}
		\caption{$M=2$}
		\label{fig:learning2}
	\end{subfigure}%
	\begin{subfigure}{0.24\textwidth}
		\centering
		\includegraphics[width=\linewidth]{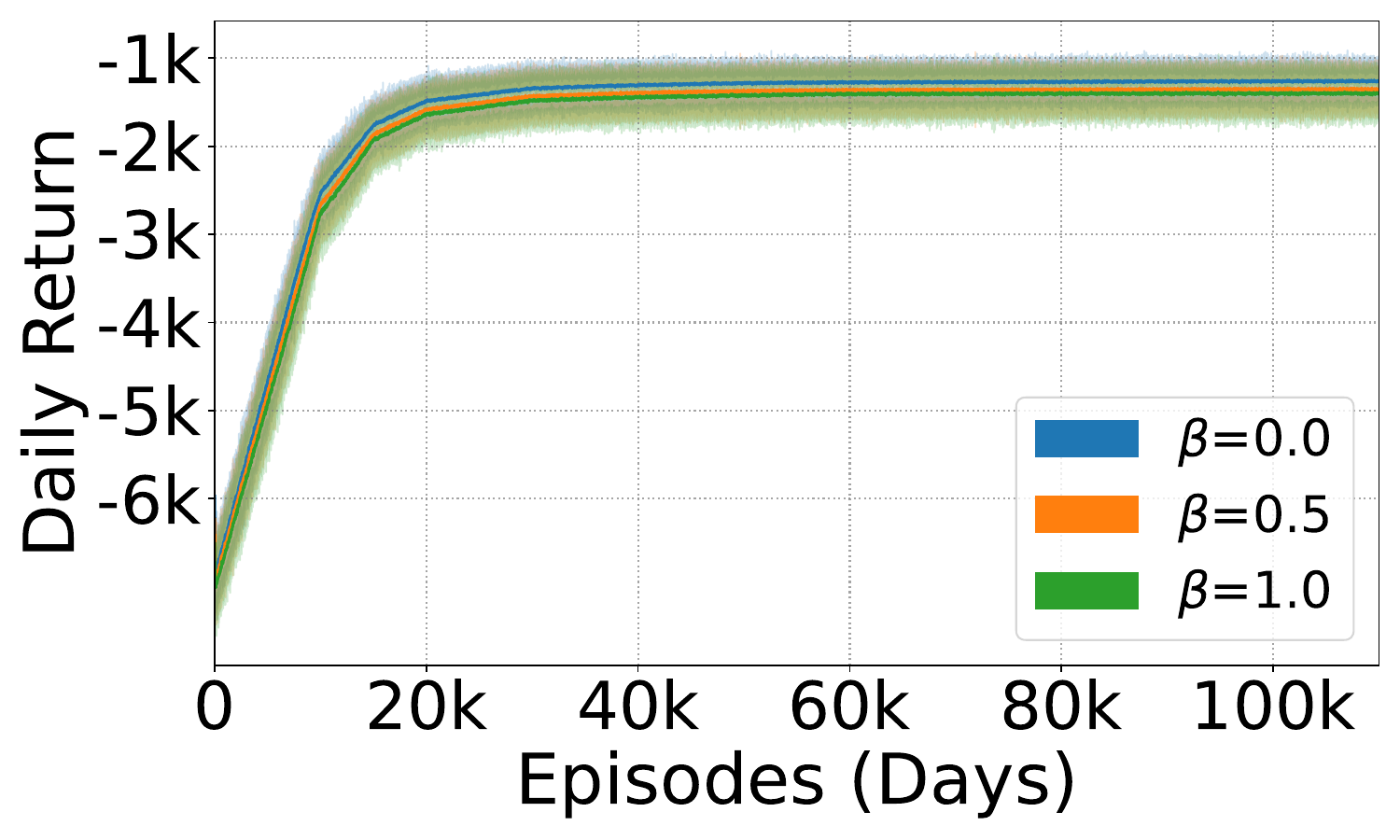}
		\caption{$M=5$}
		\label{fig:learning5}
	\end{subfigure}%
	\caption{Convergence behavior of the algorithm over training (mean $\pm 1.96$ standard dev.) for representative values of $\beta$.}
	\label{fig:learning}
\end{figure}
\begin{figure*}[htbp]
	\centering
	\begin{subfigure}{0.24\textwidth}
		\centering
		\includegraphics[width=\linewidth]{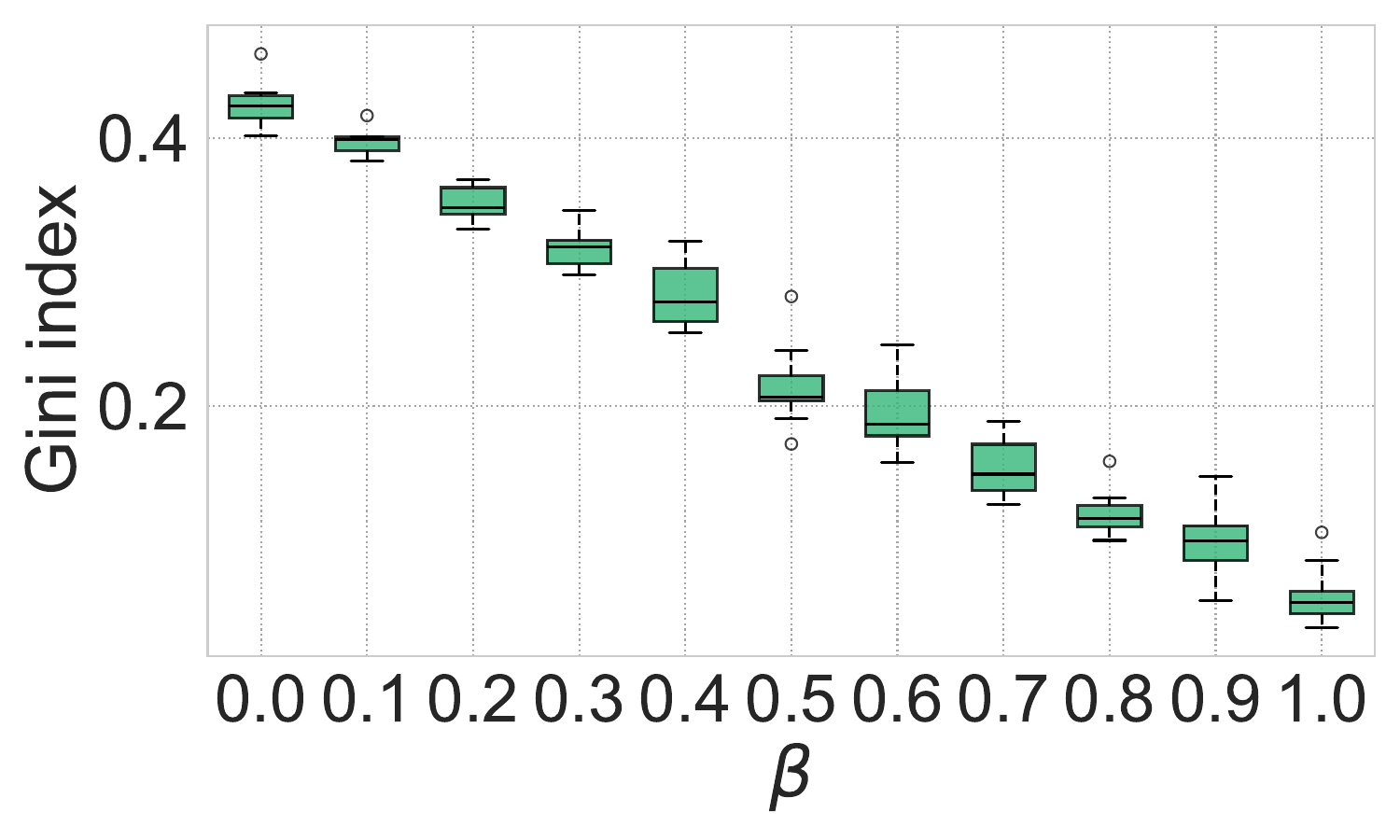}
		\caption{
			$g(x)$ vs. to $\beta$ ($M=5$)
		}
		\label{fig:boxgini}
	\end{subfigure}%
	\begin{subfigure}{0.24\textwidth}
		\centering
		\includegraphics[width=\linewidth]{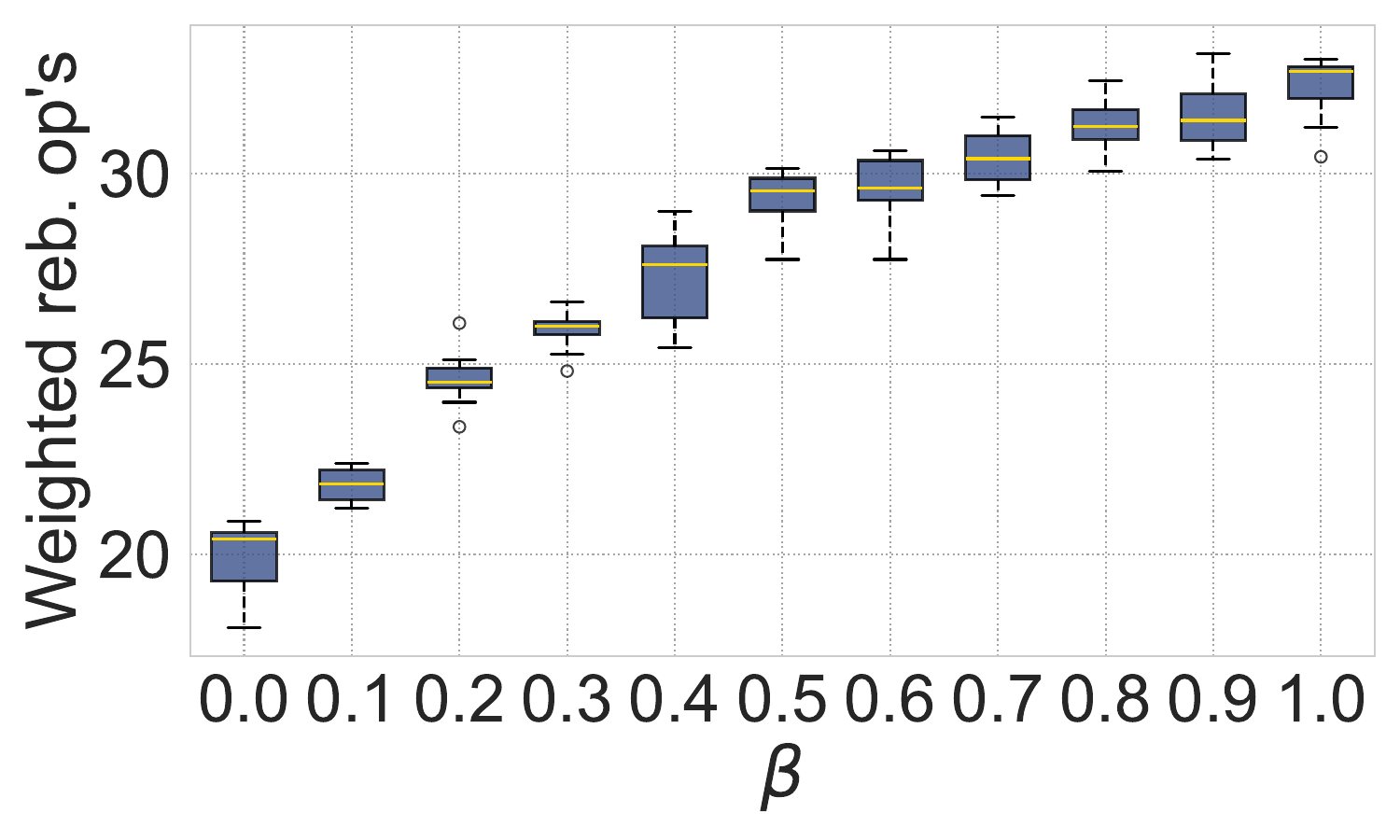}
		\caption{
			$\mathcal{C}_1$ vs. $\beta$ ($M=5$)
		}
		\label{fig:boxreb}
	\end{subfigure}%
	\begin{subfigure}{0.24\textwidth}
		\centering
		\includegraphics[width=\linewidth]{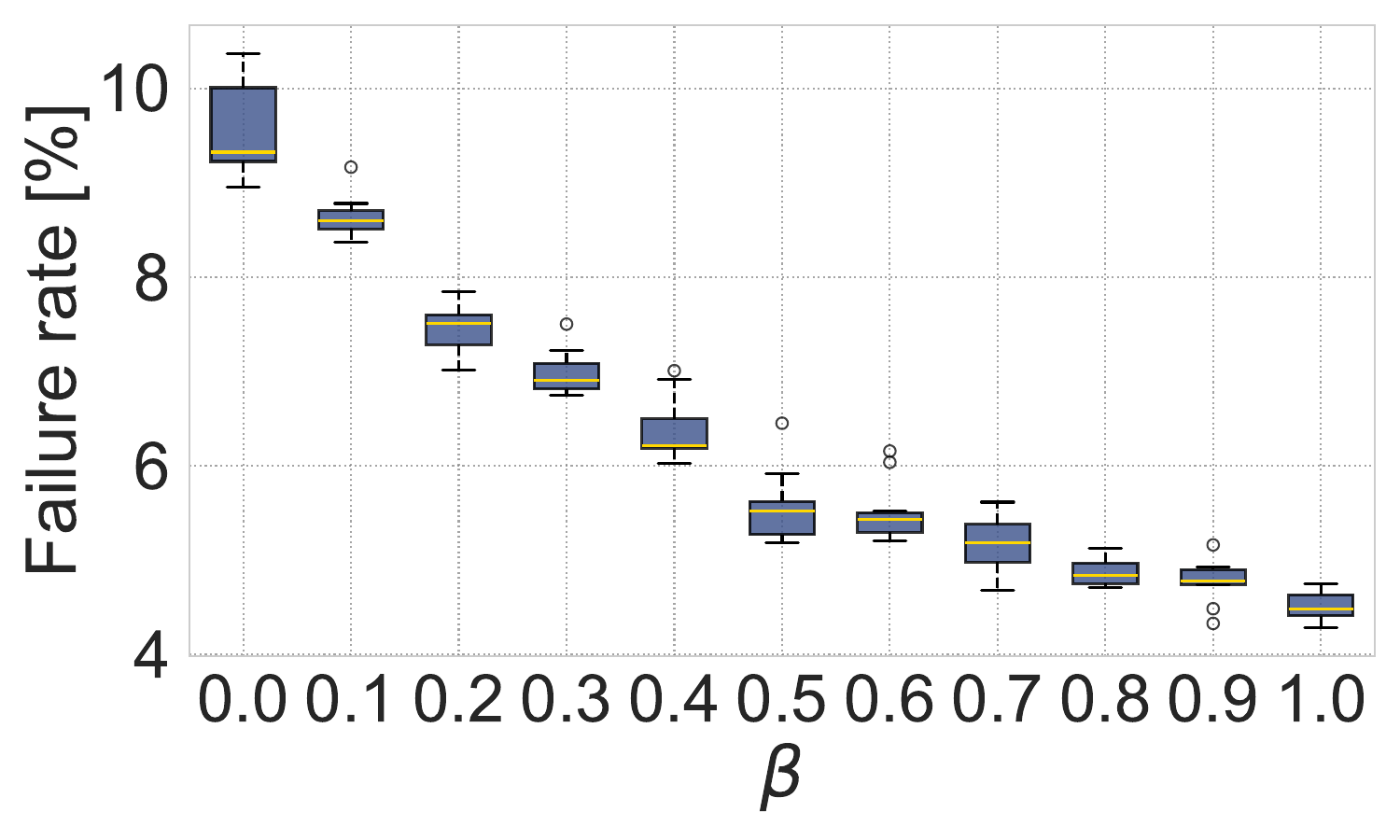}
		\caption{
			$\mathcal{C}_2$ vs. $\beta$ ($M=5$)
		}
		\label{fig:boxfails}
	\end{subfigure}%
	\begin{subfigure}{0.24\textwidth}
		\centering
		\includegraphics[width=\linewidth]{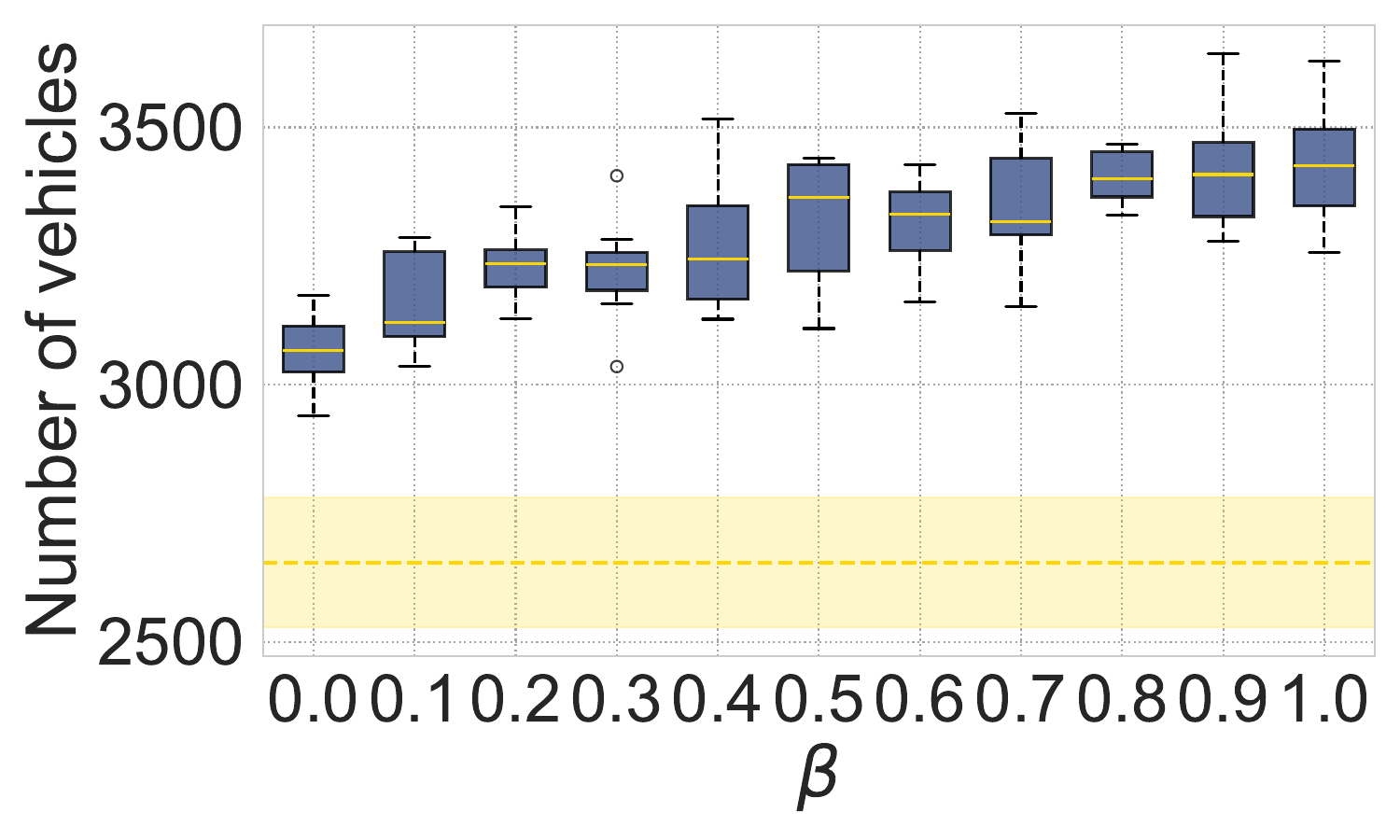}
		\caption{
			$\mathcal{C}_3$ vs. $\beta$ ($M=5$)
		}
		\label{fig:boxbikes}
	\end{subfigure}
	\caption{Distributions over 10 Monte Carlo runs of fairness and cost-related performance metrics as a function of $\beta$. (d): The yellow dashed line and area indicate resp. average and $97.5$th percentile of the distribution of the initial number of vehicles. 
	}
	\label{fig:boxplots}
\end{figure*}
\subsection{Pareto fronts for the proposed approach}

To examine the trade-off between global service cost and fairness degree of the proposed approach we have determined the Pareto front for each of the four scenarios. Upon training and evaluating the algorithm ten times across different seeds for each value of $\beta\in[0,1]$, with step-size $0.1$, the average global service cost and Gini index fairness indicator 
have been respectively compared on the x and y axes of the Pareto diagrams in Figure \ref{fig:paretos}.
Specifically, we have considered as global service cost the linear combination $\mathcal{C} := \sum_{k=1}^3 \omega_k \mathcal{C}_k$ of three sources of expenses for the service provider, where\footnote{Quantities $\text{reb}_t, f_{t,i}$, $\mu_{t,i}$ and $s_{t,i}^v$  were defined in Section~\ref{sec:methodology},
	while $\omega_1:=1, \omega_2:=10, \omega_3:=0.01$ are assumed to be given scaling factors.}
\begin{align}
	\mathcal{C}_1 &:= \text{\small$E^{-1}$} \text{\Large $\Sigma$}_{t=1}^{E} \mathrm{reb}_{t}, \quad &&\text{with }  \mathrm{reb}_t \text{ as in \eqref{eq:rebt},}\label{eq:C1}\\
	\mathcal{C}_2 &:= \text{\small$E^{-1}$} \text{\Large $\Sigma$}_{t=1}^{E} \mathrm{fail}_t, \quad &&\text{with }  \mathrm{fail}_t := \text{\Large $\Sigma$}_{i\in \mathcal{V}} \tfrac{f_{t,i}}{\mu_{t,i}}, \label{eq:C2}\\
	\mathcal{C}_3 &:= \text{\small$E^{-1}$} \text{\Large $\Sigma$}_{t=1}^{E} \mathrm{veh}_t, \quad &&\text{with }  \mathrm{veh}_t := \text{\Large $\Sigma$}_{i\in \mathcal{V}} s_{t,i}^v, \label{eq:C3}
\end{align}
respectively denote the number of rebalancing operations, the overall service failure rate and total number of vehicles. This definition is slightly different from the reward function~\eqref{eq:global_reward}, as it considers the failure rate instead of the total number of failures and the total number of vehicles without any offsets: this can better reflect the actual costs and income of an \ac{MSS} operator, while it is less effective as a reward function. Also, for completeness, Figure \ref{fig:learning} reports the convergence behavior of the proposed approach in the cases $M=2,5$. Finally, the algorithm takes on average between 7.2 and 16.7 minutes to converge on an Intel Core i7-6700 CPU, depending on $M$.

The Pareto-efficient solutions composing the frontier suggest valid choices of implementation, depending both on the desired level of fairness and the costs that the service provider is willing to bear. 
In Figures \ref{fig:pareto2} and \ref{fig:pareto3}, we can also note that, as $\beta$ approaches $1$, the solutions are not Pareto-efficient. This is expected because, as $\beta$ increases, decisions tend to become unfair towards the most central areas, since they are almost ignored when performing rebalancing operations (see (\ref{eq:our-reward})). It can be shown that this phenomenon also occurs for the scenarios $M\in\{4, 5\}$ for $\beta>1$.

Lastly, choosing $M=5$ and $\beta = 1$ leads to the highest ratio $\rho$ between maximum Gini index decrease ($-86.3\%$)
and minimum increase for $\mathcal{C}$ ($+30.0\%$)
with respect to applying no equity adjustment, i.e., with $\beta=0$.

\subsection{Fairness and costs trends for the five-category scenario}
As said above, the scenario~$M=5$ is explored more in detail in order to examine the distributions of both the fairness indicator and the three cost terms \eqref{eq:C1}, \eqref{eq:C2}, \eqref{eq:C3} encountered by the service provider as $\beta$ varies. 
It can be appreciated that the monotonic trend in the Pareto front of Figure~\ref{fig:pareto5} is, as expected, consistent with the decreasing curve of the Gini index depicted in Figure~\ref{fig:boxgini} and the increasing curves of costs $\mathcal{C}_1$ and $\mathcal{C}_3$ respectively shown in Figures~\ref{fig:boxreb},~\ref{fig:boxbikes}. 
On the other hand, as illustrated in Figure~\ref{fig:boxfails}, the decrease of $\mathcal{C}_2$ due to better service in disadvantaged neighborhoods is not enough to compensate for the higher costs needed to perform rebalancing operations (Figure~\ref{fig:boxreb}) and maintain more vehicles in the network (Figure~\ref{fig:boxbikes}).

\section{Conclusions and future directions}
\label{sec:conclusions}

This study focuses on MSS rebalancing with an emphasis on spatial fairness. 
A novel RL approach resting on the network component categorization as different city areas has been designed and tested according to the selected system performance, which is based on total number of service failures, cost of all vehicles, cost of rebalancing actions and the Gini index for vehicle accessibility. 
Numerical results lead to balanced solutions characterized by
Pareto fronts~showing a sharp trade-off between overall cost and spatial fairness.

To the best of our knowledge, this work is the first to explore this trade-off in \ac{MSS} management and not just planning, considering a Reinforcement Learning perspective. We therefore believe that this work will lead to several extensions in the future: for example, we are planning to include time-varying demands and to consider correlations between arrival/departure processes in future formalizations.


\bibliographystyle{IEEEtran}
\bibliography{references}

\begin{thebibliography}{10}
\providecommand{\url}[1]{#1}
\csname url@samestyle\endcsname
\providecommand{\newblock}{\relax}
\providecommand{\bibinfo}[2]{#2}
\providecommand{\BIBentrySTDinterwordspacing}{\spaceskip=0pt\relax}
\providecommand{\BIBentryALTinterwordstretchfactor}{4}
\providecommand{\BIBentryALTinterwordspacing}{\spaceskip=\fontdimen2\font plus
\BIBentryALTinterwordstretchfactor\fontdimen3\font minus
  \fontdimen4\font\relax}
\providecommand{\BIBforeignlanguage}[2]{{%
\expandafter\ifx\csname l@#1\endcsname\relax
\typeout{** WARNING: IEEEtran.bst: No hyphenation pattern has been}%
\typeout{** loaded for the language `#1'. Using the pattern for}%
\typeout{** the default language instead.}%
\else
\language=\csname l@#1\endcsname
\fi
#2}}
\providecommand{\BIBdecl}{\relax}
\BIBdecl

\bibitem{alleyne2023control}
A.~Alleyne \emph{et~al.}, ``Control for societal-scale challenges: Road map
  2030,'' in \emph{WS on Control for Soc.-Scale Chall.}\hskip 1em plus 0.5em
  minus 0.4em\relax IEEE CSS, 2023.

\bibitem{villa2023fair}
E.~Villa \emph{et~al.}, ``Fair-{MPC}: a control-oriented framework for socially
  just decision-making,'' 2023.

\bibitem{cheng2021role}
L.~Cheng \emph{et~al.}, ``The role of bike sharing in promoting transport
  resilience,'' \emph{Networks and Spatial Economics}, pp. 1--19, 2021.

\bibitem{dell2014bike}
M.~Dell'Amico \emph{et~al.}, ``The bike sharing rebalancing problem:
  Mathematical formulations and benchmark instances,'' \emph{Omega}, vol.~45,
  pp. 7--19, 2014.

\bibitem{chiariotti2020bike}
F.~Chiariotti \emph{et~al.}, ``A bike-sharing optimization framework combining
  dynamic rebalancing and user incentives,'' \emph{ACM Trans. Auton. and
  Adaptive Systems (TAAS)}, vol.~14, no.~3, pp. 1--30, 2020.

\bibitem{guan2024shared}
X.~Guan \emph{et~al.}, ``Shared micro-mobility and transport equity: A case
  study of three {European} countries,'' \emph{Cities}, vol. 153, p. 105298,
  2024.

\bibitem{hosford2018public}
K.~Hosford \emph{et~al.}, ``Who are public bicycle share programs serving? an
  evaluation of the equity of spatial access to bicycle share service areas in
  {Canadian} cities,'' \emph{T.R.R.}, vol. 2672, no.~36, pp. 42--50, 2018.

\bibitem{brown2021docked}
S.~Meng \emph{et~al.}, ``Docked vs. dockless equity: Comparing three
  micromobility service geographies,'' \emph{J.T.G.}, vol.~96, p. 103185, 2021.

\bibitem{chen2020dockless}
Z.~Chen \emph{et~al.}, ``Dockless bike-sharing systems: what are the
  implications?'' \emph{Transport Reviews}, vol.~40, no.~3, pp. 333--353, 2020.

\bibitem{caton2024fairness}
S.~Caton \emph{et~al.}, ``Fairness in machine learning: A survey,'' \emph{ACM
  Computing Surveys}, vol.~56, no.~7, pp. 1--38, 2024.

\bibitem{soja2009city}
E.~Soja, ``The city and spatial justice,'' \emph{Justice Spatiale/Spatial
  Justice}, vol.~1, no.~1, pp. 1--5, 2009.

\bibitem{ryden1994parameter}
T.~Ryd{\'e}n, ``Parameter estimation for markov modulated poisson processes,''
  \emph{Stochastic Models}, vol.~10, no.~4, pp. 795--829, 1994.

\bibitem{andronov2011markov}
A.~M. Andronov, ``Markov-modulated birth-death processes,'' \emph{Automatic
  Control and Computer Sciences}, vol.~45, pp. 123--132, 2011.

\bibitem{skellam1946frequency}
J.~G. Skellam, ``The frequency distribution of the difference between two
  {Poisson} variates belonging to different populations,'' \emph{J. the Royal
  Stat. Soc. Series A: Stat. in Soc.}, vol. 109, no.~3, pp. 296--296, 1946.

\bibitem{abramowitz1968handbook}
M.~Abramowitz \emph{et~al.}, \emph{Handbook of mathematical functions with
  formulas, graphs, and mathematical tables}.\hskip 1em plus 0.5em minus
  0.4em\relax US Government printing office, 1968, vol.~55.

\bibitem{weinreich2023automatic}
N.~A. Weinreich \emph{et~al.}, ``Automatic bike sharing system planning from
  urban environment features,'' \emph{Transportmetrica B: Transport Dynamics},
  vol.~11, no.~1, p. 2226347, 2023.

\bibitem{sutton2018reinforcement}
R.~S. Sutton and A.~G. Barto, \emph{Reinforcement learning: An
  introduction}.\hskip 1em plus 0.5em minus 0.4em\relax MIT press, 2018.

\bibitem{becker2004solving}
R.~Becker \emph{et~al.}, ``Solving transition independent decentralized
  {Markov} decision processes,'' \emph{J. AI Res.}, vol.~22, pp. 423--455,
  2004.

\bibitem{szepesvari1997asymptotic}
C.~Szepesv{\'a}ri, ``The asymptotic convergence-rate of q-learning,'' in
  \emph{Advances in Neural Information Processing Systems}, vol.~10, 1997.

\bibitem{barocas-hardt-narayanan}
S.~Barocas \emph{et~al.}, \emph{Fairness and Machine Learning: Limitations and
  Opportunities}.\hskip 1em plus 0.5em minus 0.4em\relax MIT Press, 2023.

\bibitem{du2020fairnessdeeplearningcomputational}


\bibitem{duran2020fair}
D.~Duran-Rodas \emph{et~al.}, ``How fair is the allocation of bike-sharing
  infrastructure? framework for a qualitative and quantitative spatial fairness
  assessment,'' \emph{T.R. (A)}, vol. 140, pp. 299--319, 2020.

\bibitem{su2024spatial}
L.~Su \emph{et~al.}, ``Spatial equity of micromobility systems: A comparison of
  shared e-scooters and docked bikeshare in {Washington DC},'' \emph{Transport
  Policy}, vol. 145, pp. 25--36, 2024.

\bibitem{gini1921measurement}
C.~Gini, ``Measurement of inequality of incomes,'' \emph{The Economic Journal},
  vol.~31, no. 121, pp. 124--125, 1921.

\bibitem{friedler2019comparative}
S.~A. Friedler \emph{et~al.}, ``A comparative study of fairness-enhancing
  interventions in machine learning,'' in \emph{FAT*}.\hskip 1em plus 0.5em
  minus 0.4em\relax ACM, 2019, p. 329.

\end{thebibliography}

\end{document}